\documentclass[a4paper]{article}
\usepackage{RR}
\usepackage{hyperref}

\usepackage{hyperref}
\usepackage{path}

\newcommand{\cgal}{{\sc Cgal}}
\newcommand{\leda}{{\textsc{Leda}}}
\newcommand{\core}{{\textsc{Core}}}
\newcommand{\gmp}{{\textsc{Gmp}}}
\newcommand{\Dag}{{\textsc{Dag}}}

\RRdate{August 2006}
\RRauthor{
Sylvain Pion\thanks{Projet Geometrica, INRIA Sophia-Antipolis,
                     \texttt{Sylvain.Pion@sophia.inria.fr}}%
  \and
Andreas Fabri\thanks{GeometryFactory, Grasse, France,
                     \texttt{andreas.fabri@geometryfactory.com}}%
}
\authorhead{Pion \& Fabri}
\RRtitle{A Generic Lazy Evaluation Scheme for\\
         Exact Geometric Computations}
\RRetitle{A Generic Lazy Evaluation Scheme for\\
          Exact Geometric Computations}
\titlehead{A Generic Lazy Evaluation Scheme for\\
          Exact Geometric Computations}
\RRresume{
Nous pr\'esentons une architecture g\'en\'erique pour effectuer des
calculs g\'eom\'etriques exacts et efficaces en utilisant une \'evaluation
paresseuse.  Les calculs g\'eom\'etriques exacts sont critiques pour
la robustesse des algorithmes g\'eom\'etriques.  Leur efficacit\'e est
\'egalement critique pour la plupart des applications, d'o\`u le besoin
pour repousser les calculs exacts le plus tard possible \`a l'ex\'ecution,
jusqu'au point o\`u ils sont absolument n\'ecessaires.  Notre approche
est g\'en\'erique et extensible dans le sens o\`u il est possible d'en faire
une biblioth\`eque que les utilisateurs peuvent \'etendre \`a leur
propres objets g\'eom\'etriques et primitives.  Elle fait appel
\`a des techniques comme les adaptateurs g\'en\'eriques de foncteurs,
le polymorphisme dynamique, le comptage de r\'ef\'erences pour la gestion
des graphes acycliques dirig\'es et la gestion d'exceptions pour signaler
les cas o\`u les calculs exacts sont requis.  Elle s'appuie \'egalement sur
le calcul arithm\'etique multipr\'ecision et l'arithm\'etique d'intervalles.
Nous appliquons notre approche au noyau g\'eom\'etrique de CGAL en entier.
}
\RRabstract{
We present a generic C++ design to perform efficient and exact geometric
computations using lazy evaluations.  Exact geometric computations are
critical
for the robustness of geometric algorithms.  Their efficiency is also critical
for most applications, hence the need for delaying the exact computations at
run time until they are actually needed.  Our approach is generic and
extensible in the sense that it is possible to make it a library which users
can extend to their own geometric objects or primitives.  It involves
techniques such as generic functor adaptors, dynamic polymorphism, reference
counting for the management of directed acyclic graphs and exception handling
for detecting cases where exact computations are needed.  It also relies on
multiple precision arithmetic as well as interval arithmetic.  We apply our
approach to the whole geometric kernel of CGAL.
}
\RRmotcle{g\'eom\'etrie algorithmique, calcul g\'eom\'etrique exact,
          robustesse num\'erique, arithm\'etique d'intervalles,
	  \'evaluation paresseuse, programmation g\'en\'erique, C++, CGAL}
\RRkeyword{computational geometry, exact geometric computation, numerical
	  robustness, interval arithmetic, lazy evaluation, generic
          programming, C++, CGAL}
\RRprojets{Geometrica}
\RRtheme{\THSym} 
\URSophia 
\begin{document}
\makeRR   

\section{Introduction}
\label{sec:introduction}

Non-robustness issues due to numerical approximations are well known
in geometric computations, especially in the computational geometry
literature.  The development of the \cgal\ library, a large collection
of geometric algorithms implemented in C++, expressed the need for
a generic and efficient treatment of these problems.

Typical solutions to solve these problems involve exact arithmetic
computations.  However, due to efficiency issues, good implementations
make use of arithmetic filtering techniques to benefit from the speed of
certified floating-point approximations like interval arithmetic, hence calling
the costly multiprecision routines rarely.

One efficient approach is to perform lazy exact computations at the level
of geometric objects.  It is mentioned in~\cite{p-gacg-99} and an implementation
is described in~\cite{fm-lookg-02}.  Unfortunately, this implementation does
not use the generic programming paradigm, although the approach is general.
This is exactly the novelty of this paper.

In this paper, we devise a generic design to provide the most generally
applicable methods to a large number of geometric primitives.  Our design
makes it easy to apply to the complete geometry kernel of \cgal, and is
extensible to the user's new geometric objects and geometric primitives.

Our design thus implements lazy evaluation of the exact
geometric objects.  The computation is delayed until a point where the
approximation with interval arithmetic is not precise enough to decide
safely comparisons, which may hopefully never be needed.

Section~\ref{sec:egc_cgal_kernel} describes in more detail the context and
motivation in geometric computing, as well as the basics of a generic geometric
kernel parameterized by the arithmetic, and what can be done at this level.
Then, Section~\ref{sec:design} discusses our design in detail, namely how
geometric predicates, constructions and objects are adapted.
Section~\ref{sec:extensibility} illustrates how our scheme can be applied to
the users's own geometric objects and primitives.  We then provide in
Section~\ref{sec:benchmarks} some benchmarks that confirm the benefit of our
design and implementation.  Finally, we list a few open questions related to
our design in Section~\ref{sec:open_questions}, and conclude with ideas for
future work.

\section{Exact geometric computations and the CGAL kernel}
\label{sec:egc_cgal_kernel}

\subsection{Exact Geometric Computations}

Many geometric algorithms such as convex hull computations, Delaunay
triangulations, mesh generators, are notoriously prone to robustness issues due
to the approximate nature of floating-point computations.  This is due to the
dual nature of geometric algorithms: on one side numerical data is used, such
as coordinates of points, and on the other side discrete structures are built,
such as the graph representing a mesh.

\begin{figure}[h]
\begin{center}
\includegraphics[width=6cm]{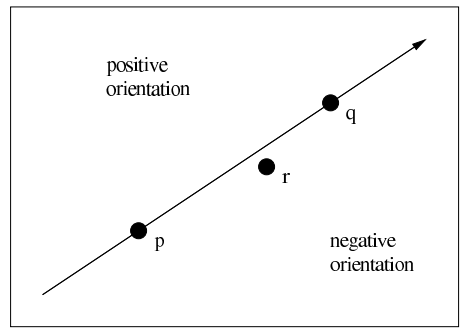}
\end{center}
\caption{The orientation predicates of 3 points in the plane.
\label{Fig::orientation}}
\end{figure}

The bridges between the numerical data and the Boolean decisions which allow
to build a discrete structure, are called the geometric
\textit{predicates}.  These are functions taking geometric objects such as points
as input and returning a Boolean or enumerated value.  Internally, these
functions typically perform comparisons of numerical values computed from
the input.  A classical example is the \texttt{orientation} predicate of three
points in the plane, which returns if the three points are doing a left turn,
a right turn, or if they are collinear (see Figure~\ref{Fig::orientation}).
Using Cartesian coordinates for the
points, the orientation is the sign (as a three-valued function: -1, 0, 1) of the
following 3-dimensional determinant which reduces to a 2-dimensional one:
$$
\begin{array}{|ccc|}
1     & 1     & 1     \\
p.x() & q.x() & r.x() \\
p.y() & q.y() & r.y() \\
\end{array}
=
\begin{array}{|cc|}
q.x() - p.x() & r.x() - p.x() \\
q.y() - p.y() & r.y() - p.y() \\
\end{array}
$$

Many predicates are built on top of signs of polynomial expressions over the
coordinates of the input points.  Evaluating such a function with
floating-point arithmetic is going to introduce roundoff errors, which can
have for consequence that the sign of the approximate value differs from
the sign of the exact value.  The impact of wrong signs on the geometric
algorithms which call the predicates can be disastrous, as for example
it can break some invariants like planarity of a graph, or make the algorithm
loop.  Didactic examples of consequences can be found
in~\cite{kmpsy-cerpg-04} as well as in the computational geometry literature.

Operations building new geometric objects, like the point at the intersection
of two lines, the circumcenter of three non-collinear points, or the midpoint
of two points, are called geometric \textit{constructions}.  We will use the
term geometric \textit{primitives} when refering to either predicates or
constructions.


In order to tackle these non-robustness issues, many solutions have been
proposed.  We focus here on the \textit{exact
geometric computation paradigm}~\cite{y-tegc-97}, as it is a general solution.
This paradigm states that, in order to ensure the correct execution of the
algorithms, it is enough that all decisions based on predicates are taken
correctly.  Concretely, this means that all comparisons of numerical values need
to be performed exactly.

A natural way to perform the exact evaluation of predicates is to
evaluate the numerical expressions using exact arithmetic.  For example,
since most computations are signs of polynomials, it is enough to use
multiprecision rational arithmetic which is provided by libraries such as
\gmp~\cite{g-ggmpa-}.  Note that exact arithmetic is also available for
all algebraic computations using libraries such as \core~\cite{klpy-clp-99} or
\leda~\cite{bms-lcrn-96}, which is useful when doing geometry over curved
objects.  This solution works well, but it tends to be very slow.

\subsection{The Geometry Kernel of CGAL}

\cgal~\cite{cgal:eb-06} is a large collection of computational geometry
algorithms.  These algorithms are parameterized by the geometry they apply to.
The geometry takes the form of a
\textit{kernel}~\cite{hhkps-aegk-01,cgal:bfghhkps-k23-06} regrouping the types of the
geometric objects such as points, segments, lines, ... as well as the basic
primitives operating on them, in the form of functors.  The \cgal\ kernel
provides over 100 predicates and 150 constructions, hence uniformity and
genericity is crucial when treating them, from a maintenance point of view.

\cgal\ provides several models of kernels.  The basic families are the template
classes \texttt{Cartesian} and \texttt{Homogeneous} which are parameterized by
the type representing the coordinates of the points.  They respectively use
Cartesian and homogeneous representations of the coordinates, and their
implementation looks as follows:
\begin{verbatim}
  template < class NT >
  struct Cartesian {
    // Geometric objects
    typedef ...          Point_2;
    typedef ...          Point_3;
    typedef ...          Segment_2;
    ...
    // Functors for predicates
    typedef ...          Compare_x_2;
    typedef ...          Orientation_2;
    ...
    // Functors for constructions
    typedef ...          Construct_midpoint_2;
    typedef ...          Construct_circumcenter_2;
    ...
  };
\end{verbatim}
These simple template models already allow to use \texttt{double} arithmetic
or multiprecision rational arithmetic for example.  \cgal\ therefore provides
a hierarchy of concepts for the \textit{number types}, which describe the
requirements for types to be pluggable into these kernels, such as addition,
multiplication, comparisons...  The functors are implemented in the following
way (here the return type of the predicate is a three-valued enumerated type,
moreover some \texttt{typename} keywords are removed for clarity):
\begin{verbatim}
  template < class Kernel >
  class Orientation_2 {
    typedef Kernel::Point_2     Point;
    typedef Kernel::FT          FT;
  public:
    typedef CGAL::Orientation   result_type;

    result_type
    operator()(Point p, Point q, Point r) const
    {
      FT det = (q.x() - p.x()) * (r.y() - p.y())
             - (r.x() - p.x()) * (q.y() - p.y());
      if (det > 0) return POSITIVE;
      if (det < 0) return NEGATIVE;
      return ZERO;
    }
  };

  template < class Kernel >
  class Construct_midpoint_2 {
    typedef Kernel::Point_2     Point;
  public:
    typedef Point               result_type;

    result_type
    operator()(Point p, Point q) const
    {
      return Point( (p.x() + q.x()) / 2,
                    (p.y() + q.y()) / 2 );
    }
  };
\end{verbatim}
As much as conversions between number types are useful, \cgal\ also provides
tools to convert geometric objects between different kernels.  We shortly
present these here as they will be refered to in the sequel.  A kernel converter is
a functor whose function operator is overloaded for each object of the source
kernel and which returns the corresponding object of the target kernel.
Such conversions may depend on the details of representation of the geometric
objects, such as homogeneous versus Cartesian representation.  \cgal\ provides
such converters parameterized by converters between number types, for
example the converter between kernels of the \texttt{Cartesian} family:
\begin{verbatim}
  template < class K1, class K2, class NT_conv =
                  Default_conv<K1::FT, K2::FT> >
  struct Cartesian_converter {
    NT_conv cv;

    K2::Point_2
    operator()(K1::Point_2 p) const
    {
      return K2::Point_2( cv(p.x()), cv(p.y()) );
    }
    ...
  };
\end{verbatim}
Related to this, \cgal\ also provides a way to find out the type of a geometric
object (say, a 3D segment) in a given kernel, given its type in another kernel
and this second kernel.  This is in practice the return type of the function
operator of the kernel converter described above.
\begin{verbatim}
  template < class O1, class K1, class K2 >
  struct Type_mapper {
    typedef ...  type;
  };
\end{verbatim}
The current implementation works by specializing on all known kernel object types
like \texttt{K1::Point\_2}, \texttt{K1::Segment\_3}.  A more extensible approach
could be sought, although this is not the main point of this paper.


\subsection{A Generic Lazy Exact Number Type}

In order to speed up the exact evaluation of predicates, people have observed
that, given that the floating-point evaluation gives the right answer in most
cases, it should be enough to add a way to detect the cases where it
can change the sign, and rely on the costly multiprecision arithmetic only in
those cases.  These techniques are usually refered to as arithmetic filtering.

\begin{figure}[h]
\begin{center}
\includegraphics[width=7cm]{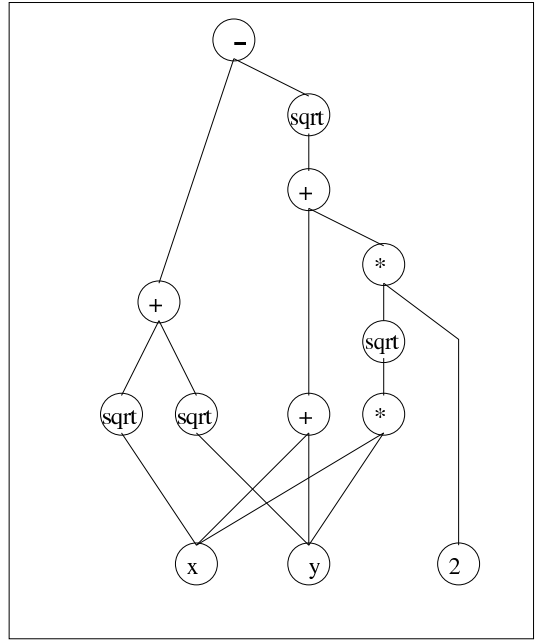}
\end{center}
\caption{Example \Dag: $\sqrt{x}+\sqrt{y}-\sqrt{x+y+2\sqrt{xy}}$.
\label{Fig::nt_dag}}
\end{figure}

There are many variants of arithmetic filters, but we are going to focus on
one which applies nicely in a generic context, and is based on interval
arithmetic~\cite{bbp-iayed-01}, a well known tool to control roundoff errors
in floating-point computations.  The idea is that we implement a new number
type which forwards its operations to an interval arithmetic type, and also
remembers the way it was constructed by storing the history of operations in
a directed acyclic graph (\Dag)~\cite{bjmm-lsicg-93}.
Figure~\ref{Fig::nt_dag} illustrates the history \Dag\ of the expression
$\sqrt{x}+\sqrt{y}-\sqrt{x+y+2\sqrt{xy}}$.

When a comparison is performed on this number type and the intervals overlap,
then the \Dag\ is used to recompute the values with an exact multiprecision
type, hence giving the exact result.  \cgal\ provides such a lazy number type
called \texttt{Lazy\_exact\_nt<NT>} parameterized by the exact type used to
perform the exact computations when needed (such as a rational number type).
Somehow, this can be seen as a wrapper on top of its template parameter, which
delays the computations until they are needed, as hopefully they won't be
needed at all.

This solution works very well.  It can however be further improved in terms of
efficiency.  Indeed we note that there are several overheads which can be
optimized.  First, a node of the \Dag\ is created for each arithmetic operation,
so it would be nice to be able to regroup them in order to diminish the number
of memory allocations as well as the memory footprint.  Second, rounding mode changes
for interval arithmetic computations are made for each arithmetic operation, so
again, it would be nice to be able to regroup them to optimize away these mode
changes.

These remark have lead to a new scheme mentioned in~\cite{p-gacg-99}, and the
description of an implementation has also been proposed in~\cite{fm-lookg-02}.
The idea is to introduce a \Dag\ at the geometric level, by considering geometric
primitives for the nodes.  The next section describes such an optimized setup.
Our design differs from the one in~\cite{fm-lookg-02} in that we followed the
generic programming paradigm and extensive use of templates to make it as
easily extensible as possible.


\section{Design of the Lazy Exact Computation Framework}
\label{sec:design}

The previously described design of lazy computation is based only on
genericity over the number type.  In this section, we make use of the
genericity at the higher level of geometric primitives, in order to
provide a more efficient solution.  We first describe how to filter
the predicates.  Then we extend the previous idea of
\texttt{Lazy\_exact\_nt} to geometric objects and constructions.

\subsection{Filtered Predicates}

Performing a filtered predicate means first evaluating the predicate
with interval arithmetic. If it fails, the predicate is evaluated again,
this time with an exact number type.  As all predicates of a \cgal\ kernel
are functors we can use the following adaptor:

{\scriptsize
\begin{verbatim}
  template <class EP, class AP, class C2E, class C2A>
  class Filtered_predicate
  {
    typedef AP    Approximate_predicate;
    typedef EP    Exact_predicate;
    typedef C2E   To_exact_converter;
    typedef C2A   To_approximate_converter;

    EP  ep;
    AP  ap;
    C2E c2e;
    C2A c2a;

  public: 

    typedef EP::result_type  result_type;

    template <class A1, class A2>
    result_type
    operator()(const A1 &a1, const A2 &a2) const
    {
      try {
        Protect_FPU_rounding P(FE_TOINFTY);
        return ap(c2a(a1), c2a(a2));
      } catch (Interval_nt_advanced::unsafe_comparison) {
        Protect_FPU_rounding P(FE_TONEAREST);
        return ep(c2e(a1), c2e(a2));
      }
    }
  };
\end{verbatim}
}

Function operators with any arity should be provided.  This is currently
done by hand up till a fixed arity, and will be replaced when variadic
templates become available in C++.

Note that \texttt{Protect\_FPU\_rounding} changes the current rounding mode
of the FPU to the one specified as argument to the constructor, and saves
the old one in the object.  Its destructor restores the saved mode, which
happens at the return of the function or when an exception is thrown.

The class {\tt Filtered\_kernel} is hence obtained from a kernel {\tt K} by
adapting all predicates of {\tt K}. This is currently done with the preprocessor.
The geometric objects as well as the constructions remain unchanged.

{\scriptsize
\begin{verbatim}
  template < class K >
  struct Filtered_kernel {

    // The various kernels
    typedef Cartesian<double>            CK;
    typedef Cartesian<Interval_nt>       AK;
    typedef Cartesian<Gmpq>              EK;

    // Kernel converters
    typedef Cartesian_converter<CK, AK>  C2A;
    typedef Cartesian_converter<CK, EK>  C2E;

    // Geometric objects
    typedef CK::Point_2                  Point_2;
    ...
    // Functors for predicates
    typedef Filtered_predicate<AK::Compare_x_2,
                               EK::Compare_x_2,
                               C2E, C2A> Compare_x_2;
    ...
  };
\end{verbatim}
}

\subsection{Lazy Exact Objects}

Performing lazy exact constructions means performing constructions with interval
approximations, and storing the sequence of construction steps. When later 
a predicate applied to these approximations cannot return a result that is guaranteed
to be correct, the sequence of construction steps is performed again, this time with 
an exact arithmetic. Now the predicate can be evaluated correctly. 

The sequence of construction steps is stored in a \Dag. Each node of the \Dag\ stores
(i) an approximation, (ii) the exact version of the function that was used to compute
the approximation, (iii) and the lazy objects that were arguments to
the function. So the outdegree of a node is the arity of the function.

\begin{figure}[h]
\begin{center}
\includegraphics{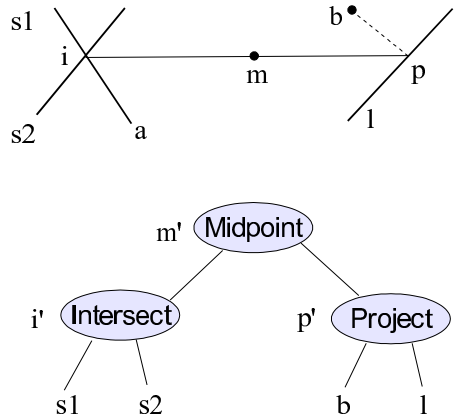}
\end{center}
\caption{The \Dag\ represents the midpoint of an intersection point
and the vertical projection of a point on a line. Testing whether $a$, $m$, and
$b$ are collinear has a good chance to trigger an exact construction.\label{Fig::tree}}
\end{figure}

The example illustrates that lazy objects can be of the same type, without being the result
of the same sequence of constructions. $a$, $m$, and $b$ are all point-ish.
Therefore we have a templated handle class, with a pointer to a node of the \Dag.
In our example, only the latter are of different types.

\begin{figure}[h]
\begin{center}
\includegraphics{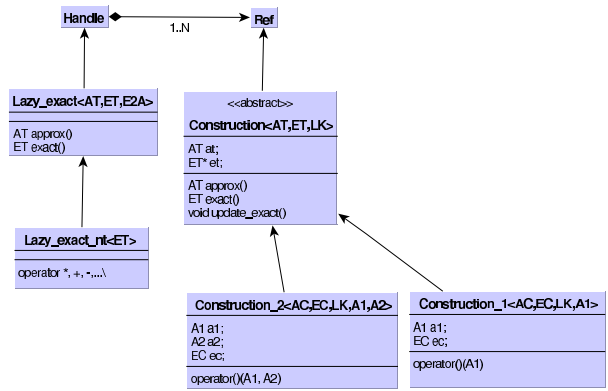} 
\end{center}
\caption{The class hierarchy for the nodes of the \Dag.
\label{Fig::uml}}
\end{figure}

We will now explain some of the classes in Figure~\ref{Fig::uml} in more detail.

{\tt Lazy\_exact} is the handle class.  It also does reference counting with a design
similar to the one described in~\cite{k-rcld-05}. It has  {\tt Lazy\_exact\_nt} as subclass,
which provides arithmetic operations.  Note that this framework handles 
arithmetic and geometric objects in a unified way. For example a distance bewteen
geometric objects yields a lazy exact number, and a lazy exact
number can become the coordinate of a point.

The class {\tt Construction} is an abstract base class. It stores
the approximation, and holds a pointer to the exact value. Initially, this
pointer is set to {\tt NULL}, and it is the virtual member function {\tt update\_exact}
which later may compute the exact value and then cache it.

The subclass {\tt Construction\_2} is used for binary functions. Similar classes
exist for the other arities. These classes store the
arguments and the exact version of the function. The arguments may be
of arbitrary types. In the case of lazy exact geometric objects or
lazy exact numbers the arguments are handles as described before.

{\scriptsize
\begin{verbatim}
  template <class AC, class EC, class LK, class A1, class A2>
  class Construction_2
    : public Construction<AC::result_type, EC::result_type, E2A>
    , private EC
  {
    typedef AC               Approximate_construction;
    typedef EC               Exact_construction;
    typedef LK::C2E          To_exact_converter;
    typedef LK::C2A          To_approximate_converter;
    typedef LK::E2A          Exact_to_approximate_converter;
    typedef AC::result_type  AT;
    typedef EC::result_type  ET;

    A1 m_a1;
    A2 m_a2;

    const EC& ec() const { return *this; }

  public:

    void
    update_exact()
    {
      this->et = new ET(ec()(C2E()(m_a1), C2E()(m_a2)));
      this->at = E2A()(*(this->et));
      // Prune lazy dag
      m_a1 = A1();
      m_a2 = A2();
    }

    Construction_2(const AC& ac, const EC& ec, 
	           const A1& a1, const A2& a2)
      : Construction<AT,ET,E2A>(ac(C2A()(a1), C2A()(a2)),
                                m_a1(a1), m_a2(a2)
    {}
  };
\end{verbatim}
}

The constructor stores the two arguments. It then takes their 
approximations and calls the approximate version of the functor.

In case the exact version of the construction is needed, this gets
computed in the {\tt update\_exact} method. It fetches the exact
versions of the arguments, which in turn may trigger their exact
computation if they are not already computed and cached. From the
exact lazy object one computes again the approximate object, as the object
computed with the approximate version of the functor has a good chance
to have accumulated more numerical error.

Finally, the \Dag\ is pruned.  As the nodes of the \Dag\ are
reference counted, some of them may get deallocated by the pruning.
Most often {\tt A1} and {\tt A2} will be lazy exact objects.
For performance reasons their default constructors generates a 
handle to a shared static node of the \Dag.

Also, we use private derivation of the exact construction {\tt EC},
instead of storing it as data member, in order to benefit from the
empty base class optimization.

The other derived classes store the leaves of the \Dag. There is a general
purpose leaf class, and more specialized ones, for example for creating
a lazy exact number from an {\tt int}.  They are there for performance
reasons.

\subsection{The Functor Adaptor}

So far we have only explained how lazy constructions are stored, but not
how new nodes of the \Dag\ are generated.

The following functor adaptor is applied to all the constructions we
want to make lazy. It has function operators for other arities.

{\scriptsize
\begin{verbatim}
  template <class LK, class AC, class EC>
  class Construct
  {
    typedef LK                  Lazy_kernel; 
    typedef AC                  Approximate_construction;
    typedef EC                  Exact_construction;
    typedef LK::AK              AK;
    typedef LK::EK              EK;
    typedef EK::FT              EFT;
    typedef LK::E2A             E2A;
    typedef LK::C2E             C2E;
    typedef AC::result_type     AT;
    typedef EC::result_type     ET;
    typedef Lazy_exact<AT, ET, E2A> Handle;

    AC ac;
    EC ec;

  public:

    typedef Type_mapper<AT,AK,LK>::type result_type;

    template <class A1, class A2>
    result_type
    operator()(const A1& a1, const A2& a2) const
    {
      try {
        Protect_FPU_rounding P(FE_TOINFTY);
        return Handle(new Construction_2<AC, EC, LK, A1, A2>
                                        (ac, ec, a1, a2));
      } catch (Interval_nt_advanced::unsafe_comparison) {
        Protect_FPU_rounding P(FE_TONEAREST);
        return Handle(new Construction_0<AT,ET,LK>
                                 (ec(C2E()(a1), C2E()(a2))));
      }
    }
  };
\end{verbatim}
}

The functor first tries to construct a new node of the \Dag. If
inside the approximate version of the construction an exception is
thrown, we perform the exact version of the construction, and only
create a leaf node for the \Dag.

\subsection{Special-Case Handling}

The generic functor adaptor works out of the box for all functors that
return lazy exact geometric objects or a lazy exact number.

Functors returning objects which are not made lazy are an easy to
handle exception. An example in {\sc Cgal} is the functor that
computes the bounding box with {\tt double} coordinates. As the
intervals of the coordinates of the approximate geometric object are
already 1-dimensional bounding boxes we never have to recur to the
exact geometric object. The functor adaptor is trivial. 

Some functors of {\sc Cgal} kernels return a polymorphic object.  For
example, the intersection of two segments may be empty, or a point, or a
segment. In order not to have a base class for all geometric classes,
{\sc Cgal} offers a class {\tt Object}\footnote{The {\tt Object} class 
is comparable to {\tt boost::any}.} which
is capable of storing typed objects.  The problem we have to solve is
that the lazy exact functor must not return a lazy exact {\tt
Object}, but instead must return an {\tt Object} holding a lazy
geometric object. This is solved by looping over all {\sc Cgal}
kernel types, to try to cast, and if it works to construct the
lazy geometric object and put it in an {\tt Object} again.

Less trivial cases are functors which pass results of a computation
back to reference parameters, or which write into output iterators.
They need a special functor as well as special {\tt Construction}
classes. It is not hard to write them, but the problem is that
they must be dispatched by hand, as we have no means of introspection.
One solution would be to introduce functor categories.

\subsection{The Lazy Exact Kernel}

We are ready to put all pieces together, by defining a new kernel
which has an approximate and an exact kernel as template parameters.

{\scriptsize
\begin{verbatim}
  template < class AK, class EK >
  struct Lazy_kernel {

    // Kernel converters
    typedef Lazy_kernel<AK, EK>          LK;
    typedef Approx_converter<LK, AK>     C2A;
    typedef Exact_converter<LK, EK>      C2E;
    typedef Cartesian_converter<EK, AK>  E2A;

    // Geometric objects
    typedef Lazy_exact<AK::Point_2,   EK::Point_2>   Point_2;
    typedef Lazy_exact<AK::Segment_2, EK::Segment_2> Segment_2;
    
    // Functors for predicates
    typedef Filtered_predicate<EK::Compare_x_2, AK::Compare_x_2,
                               C2E, C2A> Compare_x_2;
    ...
    
    // Functors for constructions
    typedef Lazy_construct<LK, AK::Construct_midpoint_2,
                           EK::Construct_midpoint_2>
            Construct_midpoint_2;
    ...

    typedef Lazy_Construct_returning_object<LK, AK::Intersection_2,
                                            EK::Intersection_2>
            Intersection_2;
  };
\end{verbatim}
}

In the current implementation we use the preprocessor to generate
the typedefs from a list of types, and we use the Boost {\sc Mpl}
library for dispatching the special cases.  \texttt{Approx\_converter}
simply fetches the stored approximate object.  Similarly
\texttt{Exact\_converter} fetches the exact approximate object,
possibly triggering its computation.



\section{Extensibility}
\label{sec:extensibility}

We have to distinguish between different levels of extensibility.

When \cgal\ kernels get extended by geometric objects and constructions 
this needs changes in the lazy construction framework if the new constructions
have ``new'' interfaces, e.g., two output iterators, followed by
two reference parameters to return a result.  This would need a 
new node type for the \Dag, a new functor, and hard wired dispatching
in the lazy kernel. Otherwise there is nothing to do.

When the \cgal\ user wants to extend the lazy kernel with his
own geometric objects and constructions he first has to add them
to the kernel that gets lazified, as described in~\cite{hhkps-aegk-01}.
Then, what we stated in the previous paragraph applies.

The {\tt Curved\_kernel} and the {\tt Lazy\_curved\_kernel} of \cgal\
which provide primitives on circles and circular
arcs~\cite{cgal:pt-cc2-06,ekptt-tock-04}, are examples for both.

\section{Benchmarks}
\label{sec:benchmarks}

We now run a simple benchmark that illustrates the benefit of our techniques.
We compare the running time and memory consumption of various kernel choices with
the following algorithm:
\begin{itemize}
\item generate 2000 pairs of 2D points with random coordinates (using \texttt{drand48()}).
\item construct 2000 segments out of these points.
\item intersect all pairs of segments among these, and store the resulting intersection
      points.
\item shuffle the resulting points
\item iterate over consecutive triplets of these points, and compute the
      \texttt{orientation} predicate of these.
\end{itemize}

Figure~\ref{Fig:bench} provides the resulting data for a choice of four different kernels:
\begin{itemize}
\item \texttt{SC$<$Gmpq$>$} stands for the simple Cartesian representation
      kernel parameterized with \texttt{Gmpq}, which is a C++ wrapper around
      the multiprecision rational number type provided by \gmp,
\item \texttt{SC$<$Lazy\_exact\_nt$<$Gmpq$>>$} uses the lazy exact evaluation mechanism at
      the arithmetic level,
\item \texttt{Lazy\_kernel<SC$<$Gmpq$>>$} is our approach for performing lazy exact
      evaluations at the geometric object level,
\item finally, \texttt{SC$<$double$>$} is the simple Cartesian representation
      kernel parameterized with \texttt{double}.  It is given for reference as it is
      not robust in all cases.  It shows what the optimal performance could be.
\end{itemize}
Benchmarks have been performed using the GNU \texttt{g++} compiler versions 3.4
and 4.1 under Linux, with the \texttt{-O2} optimization option.  The memory
consumption is the same for these two compiler versions, however timings
differ significantly.  Timings are given in seconds and memory in megabytes.

\begin{figure}[h]
\begin{center}
\begin{tabular}{|l||r|r|r|}
\hline
Kernel & time             & time              & mem \\
       & \texttt{g++} 3.4 & \texttt{g++} 4.1  &     \\
\hline\hline
\texttt{SC$<$Gmpq$>$}                     & 71   & 70   & 70  \\
\texttt{SC$<$Lazy\_exact\_nt$<$Gmpq$>>$}  & 9.4  & 7.4  & 501 \\
\texttt{Lazy\_kernel<SC$<$Gmpq$>>$}       & 4.1  & 2.8  & 64  \\
\hline\hline
\texttt{SC$<$double$>$}                   & 0.98 & 0.72 &  8.3  \\
\hline
\end{tabular}
\caption{Benchmarks comparing different kernels.
\label{Fig:bench}}
\end{center}
\end{figure}

The results show that our approach wins almost a factor of 10 on memory over
the basic lazy evaluation scheme.  It is also between 2 and 3 times faster.
However, it remains 4 times slower than the approximate floating-point
evaluation, but of course it is guaranteed for all cases.

Note that the algorithm we chose uses random data, hence it does not
produce many filter failures, so almost not exact evaluation is
performed.  Another thing worth noticing is that it uses relatively
simple 2D primitives.  More complex primitives, especially in higher
dimensions, should show more benefits to the method.  Finally,
real-world geometric applications tend to produce more combinatorial
output, hence the relative runtime cost of primitives is smaller,
so the slow down factor is lower in those cases.

\section{Open Design Questions}
\label{sec:open_questions}

Here is a list of open questions related to our framework.

The first question concerns the regrouping of expressions.  Our framework asks the user
to pass it functors specifying the level at which the regrouping of expressions is made.
In \cgal\ this is not a problem since the primary interface of the kernel towards the
geometric algorithms is a list of functors.
However it has the drawback of not being automatic.  We can think of approaches
based on expression templates~\cite{Veldhuizen95b} which would automatically
detect sequences of operations and regroup them.  Unfortunately, expression
templates are limited to single statement expressions and they tend to slow
down compilation times considerably.  Could there be a way to extend the
automatic regrouping to more than single statements?  Maybe the \texttt{auto}
keyword recently proposed for addition to the C++ language will allow to
propagate this through several statements?  Or maybe the Axiom feature part
of the proposal for concepts in C++ could be used to specify this kind of
transformation.

Another question is if similarly delayed computations are used in other areas, and
if yes, then is it possible to find out a common design, more general than the one
we propose.

\section{Conclusion and future work}
\label{sec:conclusion}
We have presented in this paper a generic framework which implements lazy exact
geometric computations, motivated by the needs for robustness and efficiency
of geometric algorithms.  This framework allows to delay the costly exact
evaluation using multiprecision arithmetic when the faster interval arithmetic
suffices.

The proposed design is easily extensible to new geometric primitives --
predicates and constructions --, as well as new geometric objects.  It is based
on a template family for representing lazy objects, as well as generic functor
adaptors which produce them.

Future work in this area will consist of various added special-case
optimizations as well as generalizations.  It is for example possible to
refine the filtering scheme by growing the precision little by little
instead of switching directly to full multiprecision computation in case
of insufficiency of precision of the intervals.  We also would like to
study possibilities of merging the \texttt{Filtered\_predicate} and
\texttt{Lazy\_construct} functor adaptors.  Possible optimizations
for specific cases also can be done, using faster schems than interval
arithmetic (so-called static filters).  Moreover, the current way of
providing a full kernel is by a list of types for the objects and functors,
which is provided through the use of the preprocessor, we will therefore try to
provide a better design on this particular point.

Finally, we plan to make our implementation part of a future release of \cgal,
whose entire geometry kernel already benefits from it.

\section{Acknowledgments}

This work was supported by the European Union's information technologies
programme Esprit, Project IST-006413 - ACS, Algorithms for Complex Shapes.

\bibliographystyle{abbrv}
\bibliography{lazy-kernel,geom,how_to_cite_cgal}

%
\newpage
\onecolumn
\appendix
\section{Benchmark code}
{\scriptsize
\begin{verbatim}
#include <CGAL/Simple_cartesian.h>
#include <CGAL/Lazy_kernel.h>
#include <CGAL/Gmpq.h>
#include <CGAL/Lazy_exact_nt.h>
#include <CGAL/intersections.h>
#include <CGAL/Timer.h>
#include <CGAL/Memory_sizer.h>
using namespace CGAL;

// Choosing a kernel:
//typedef Simple_cartesian<Gmpq>                 K;
//typedef Simple_cartesian<Lazy_exact_nt<Gmpq> > K;
//typedef Lazy_kernel<Simple_cartesian<Gmpq> >   K;
typedef Simple_cartesian<double>               K;

typedef K::Point_2    Point;
typedef K::Segment_2  Segment;

Point   random_point()   { return Point(drand48(), drand48()); }
Segment random_segment() { return Segment(random_point(), random_point()); }

int main() {
  int loops = 2000, init_mem = Memory_sizer().virtual_size();
  Timer t; t.start();

  std::cout << "Generating initial random segments: " << loops << std::endl;
  std::vector<Segment> segments;
  for (int i = 0; i < loops; ++i)
    segments.push_back(random_segment());

  std::cout << "Counting intersections [brute force algorithm]: " << std::flush;
  std::vector<Point> points;
  for (int i = 0; i < loops-1; ++i)
    for (int j = i+1; j < loops; ++j) {
      Object obj = intersection(segments[i], segments[j]);
      if (const Point* pt = object_cast<Point>(&obj))
        points.push_back(*pt);
    }
  std::cout << points.size() << std::endl;

  // we shuffle the points, as consecutive points have good chance to come
  // from the same segments, hence filter failures in orientation() later...
  std::random_shuffle(points.begin(), points.end());

  std::cout << "Performing orientation tests" << std::endl;
  int negative_ort = 0, positive_ort = 0, collinear_ort = 0;
  for (int i=0; i < points.size()-2; ++i) {
    Orientation o = orientation(points[i], points[i+1], points[i+2]);
    if (o < 0)      ++negative_ort;
    else if (o > 0) ++positive_ort;
    else            ++collinear_ort;
  }
  std::cout << "orientation results : (-) = " << negative_ort
                              << "    (+) = " << positive_ort
                              << "    (0) = " << collinear_ort << std::endl;

  t.stop();
  std::cout << "Total time   = " << t.time() << std::endl;
  std::cout << "Total memory = " << ((Memory_sizer().virtual_size() - init_mem) >>10)
                                 << " KB" << std::endl;
}
\end{verbatim}
}

\newpage
\tableofcontents

\end{document}